\documentclass[runningheads]{llncs}
\usepackage[T1]{fontenc}
\usepackage{graphicx}
\usepackage{xcolor}
\usepackage{array}
\usepackage{subcaption} 

\begin{document}
\title{OnDiscuss: An Epistemic Network Analysis Learning Analytics Visualization Tool for Evaluating Asynchronous Online Discussions}
\titlerunning{ENA LA Visualization Tool for Asynchronous Online Discussions}
%
\author{Yanye Luther\inst{1}\orcidID{0000-0002-0995-3356} \and
Marcia Moraes\inst{1}\orcidID{0000-0002-9652-3011} \and 
Sudipto Ghosh\inst{1}\orcidID{0000-0001-6000-9646} \and 
James Folkestad\inst{2}\orcidID{0000-0003-0301-8364}
\email{\{yanye.luther, marcia.moraes, sudipto.ghosh,james.folkestad\}@colostate.edu}}
\authorrunning{Y. Luther et al.}
\institute{Department of Computer Science\textsuperscript{1}, School of Education\textsuperscript{2}, Colorado State University, Fort Collins CO 80523, USA}

\maketitle              
\begin{abstract}
Asynchronous online discussions are common assignments in both hybrid and online courses to promote critical thinking and collaboration among students. However, the evaluation of these assignments can require considerable time and effort from instructors. We created OnDiscuss, a learning analytics visualization tool for instructors that utilizes text mining algorithms and Epistemic Network Analysis (ENA) to generate visualizations of student discussion data. Text mining is used to generate an initial codebook for the instructor as well as automatically code the data. This tool allows instructors to edit their codebook and then dynamically view the resulting ENA networks for the entire class and individual students. Through empirical investigation, we assess this tool's effectiveness to help instructors in analyzing asynchronous online discussion assignments.

\end{abstract}

\section{Introduction}
Social interactions play an important role in learning process. According to the social learning theory developed by Vygotsky~\cite{vygotsky1978}, learning takes place primarily in social and cultural settings where students interact with their peers, teachers, and parents as active participants in the creation of their own knowledge. Interpersonal interactions and discussions are a fundamental part of teaching and learning in this context~\cite{prawat1992teachers,davis2017learning}. Therefore, asynchronous online discussions (AOD) are widespread in hybrid and online courses~\cite{aloni2018research,fehrman2021systematic}. Some of the benefits of AOD are a deeper understanding of course material~\cite{decker2016graduate}, more effective communication with group members~\cite{decker2016graduate}, improvements in critical thinking and writing skills~\cite{aloni2018research}, and increased student performance in meeting learning outcomes~\cite{alzahrani2017effect}. Despite these benefits, instructors reported struggling to assess students’ contributions in forum activities due to difficulties in following the discussions, the lack of specific reports related to the subjects discussed, the students’ contributions to those subjects, and the lack of visualizations to convey messages in a graphical format~\cite{delima2019expect}.

In this paper, we present OnDiscuss, a learning analytics (LA) visualization tool to support instructors in evaluating asynchronous online discussions. The tool utilizes both text mining and Epistemic Network Analysis (ENA). Latent Dirichlet Allocation (LDA)~\cite{blei2003latent} is used to perform automatic topic extraction from the discussion data to create an initial codebook based on the findings of~\cite{saravani2022automated}. Bigrams and trigrams are incorporated in LDA and topics are extracted over the entire discussion thus using an infinite stanza window~\cite{saravani2022automated}. The use of LDA with the intervention and addition of the instructor's own keywords has demonstrated significant potential to assist instructors in evaluating discussion based assignments~\cite{moraes2023combining}. However, in~\cite{moraes2023combining} the ENA visualizations were only demonstrated to a single instructor, requiring manual presentation. This current study investigates the potential for broader applications by exploring how different instructors utilize a tool that enables them to modify their own codebook and observe immediate changes in ENA models.

This paper aims to answer the following research questions:

\begin{itemize}
    \item \textbf{RQ1.} Are ENA visualizations of asynchronous online discussion data helpful for instructors who are novices in ENA?
    \item \textbf{RQ2.} Do the ENA visualizations have the potential to reduce the time and effort spent assessing asynchronous online discussions for instructors of any experience level with ENA?
\end{itemize}

\section{Related Works}
ENA has been used in several works related with collaborative learning in asynchronous online discussions~\cite{rolim2019network,gavsevic2019sens,swiecki2020isens,scianna2023sseen}. Rolim et al. \cite{rolim2019network} used ENA to provide insights on the relationship between social and cognitive presence in asynchronous online discussions. Gašević et al. \cite{gavsevic2019sens} proposed the use of social epistemic network signature (SENS), which combines ENA and Social Network Analysis (SNA) to analyze collaborative learning. Swiecki and Shaffer \cite{swiecki2020isens} extended SENS and proposed the integrated social-epistemic network signature (iSENS), an approach that provides the simultaneous investigation of cognitive and social connections in collaborative learning. Scianna and Kaliisa~\cite{scianna2023sseen} proposed an analysis workflow and visualization method called Social Sentiment Embedded Epistemic Networks (SSEEN) to consider how sentiment manifests and explore its usefulness in understanding student interactions in asynchronous online discussions.

Besides its use in collaborative learning analysis, ENA has been used to visualize many aspects of learning and education~\cite{herder2018supporting,fernandeznieto2021modelling}. Fougt et al.~\cite{fougt2018epistemic} analyzed instructors' ability to assess student papers and considered a different range of number of topics in order to capture different levels of complexity of the ENA models. Visually, ENA had the potential to indicate the quality of the student assignment. The instructors reported it can be difficult to choose the correct number of codes and keywords and that it should be left to the instructor to make that informed decision. Vega and Irgens~\cite{vega2022constructing} introduced participatory quantitative ethnography (QE) which includes participants in co-construction and co-interpretation of ENA models. This work demonstrates the deeper analysis that arises from interpretations of data by modifying codes, adding connections, and reacting to codes. 

Unlike previous works that used ENA in collaborative learning analysis, this work combines LDA and ENA in a tool that: builds an initial codebook for instructor, automatically codes the data of asynchronous online discussions using that initial codebook, and generates individual ENA visualizations of student's content connections. In addition, past works do not have the ability to modify the codebook and thus instantly update the ENA model on their own without any researcher intervention.

\section{Methodology}

\subsection{Participant Selection}
In order to answer the research questions, we selected two instructors; one instructor that did not have previous experience with ENA and one instructor with previous experience with ENA to evaluate their respective asynchronous online discussion with the aid of OnDiscuss. Both instructors taught at the same Research 1 land-grant university. The instructor without experience in ENA is referred to as Instructor A, the novice, and the instructor with experience is referred to as Instructor B, the expert. Instructor A taught a graduate level Computer Science course with discussion discourse amongst small groups of 5-6 students with about 2-3 groups in the entire class. Instructor B taught a graduate level Education course with discourse amongst all 20 students in the class.

The courses that were chosen for the experiment had to meet the following criteria: the course must have asynchronous online discussion assignments; those assignments must require students to assimilate knowledge and synthesize concepts into a coherent discussion post rather; and the discussions must be entirely raw text since OnDiscuss can't parse potential keywords in images, videos, links to websites, etc.

\subsection{OnDiscuss Description and Functionality}
OnDiscuss is a learning analytics visualization tool for instructors that utilizes text mining algorithms and Epistemic Network Analysis (ENA) to generate visualizations of student discussion data. As highlighted by reviewers in~\cite{moraes2021using}, the process of establishing a codebook from scratch may present challenges for instructors new to ENA.
Our tool uses the same process described in~\cite{moraes2023combining} to generate five topics with ten keywords, each derived from the discussion's data to be used as an initial codebook for the instructor. The rationale for the initial number of topics and keywords can be found in~\cite{saravani2022automated}.
After the initial codebook is built, the tool uses text mining algorithms to automatically code the discussion for each post. That coded data is then represented in an ENA network. The tool uses rENA~\cite{rena} package and the following settings to build the ENA model: unit is the student ID, the conversation is the utterance from the student in their discussion post, and an infinite stanza window. That configuration enabled us to generate the connections between the codes that were to be discussed by each student. 

\begin{figure}[h]
    \centering
    \includegraphics[width=1.0\linewidth]{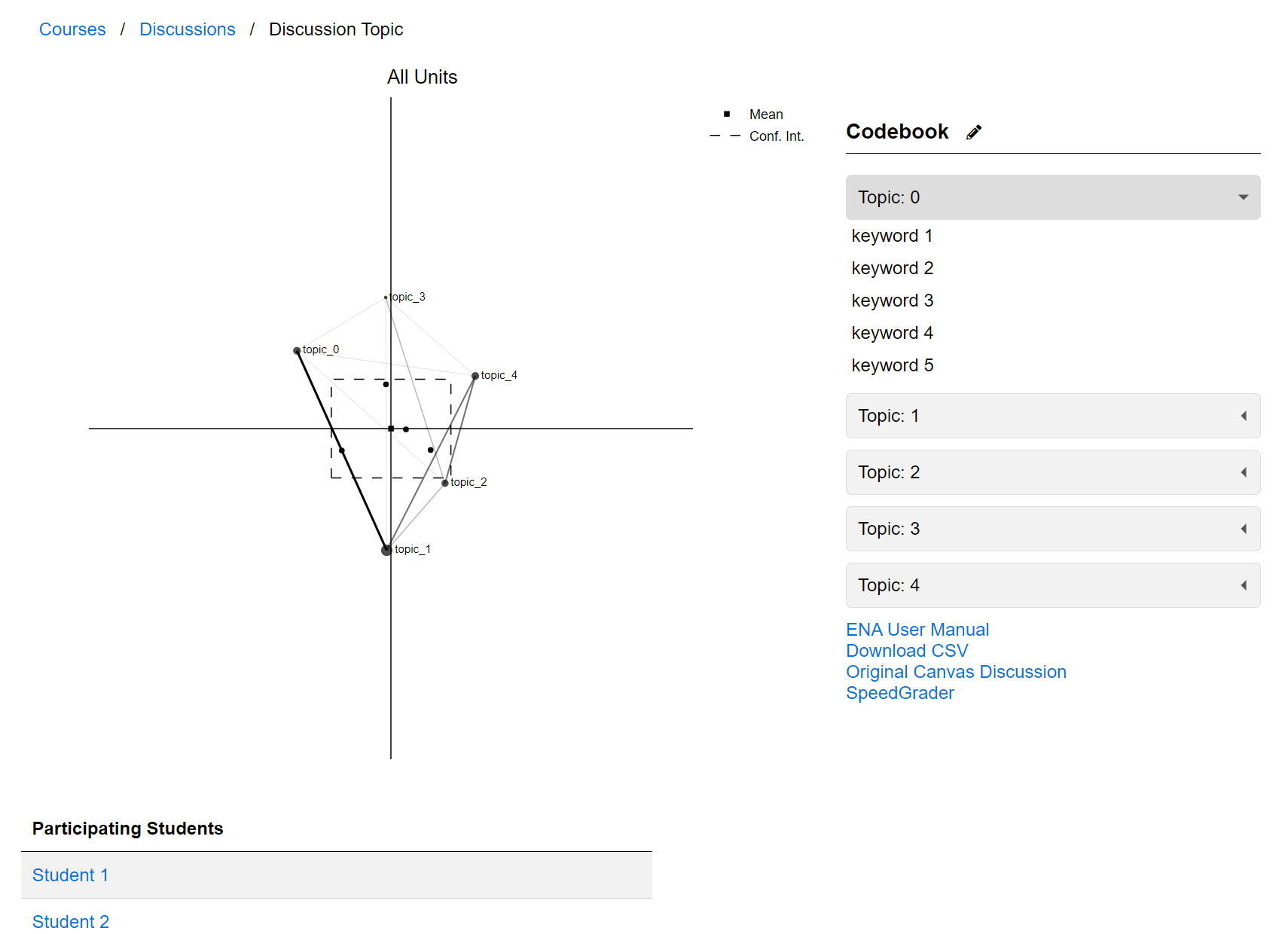}
    \caption{Example Class/Group ENA View for a Discussion Topic}
    \label{fig:tool_group}
\end{figure}

\begin{figure}[h]
    \centering
    \includegraphics[width=1.0\linewidth]{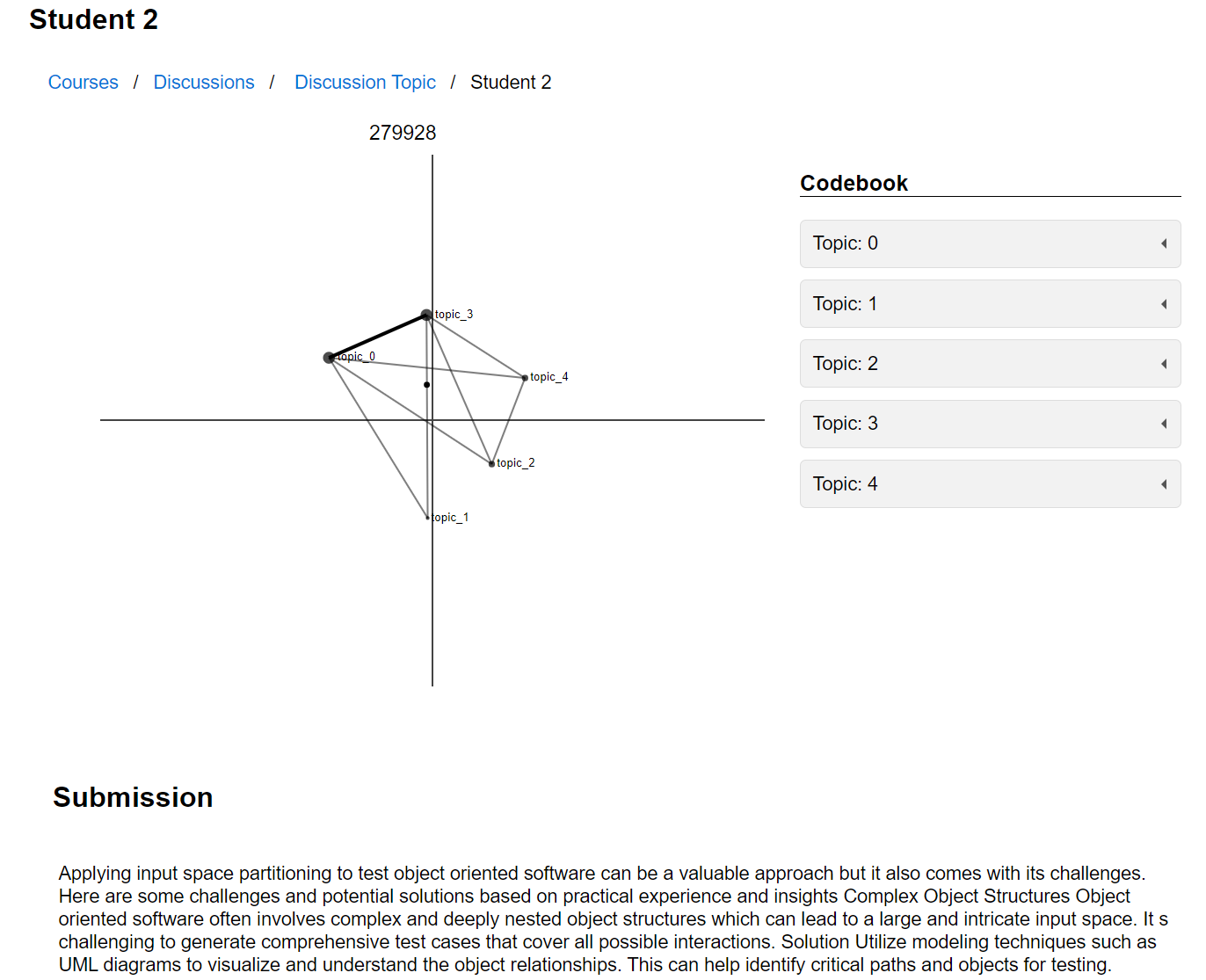}
    \caption{Example Individual ENA View for a Discussion Topic}
    \label{fig:tool_individual}
\end{figure}

OnDiscuss is integrated with Canvas Learning Management System (LMS) which allowed instructors to view a list of all discussions published to Canvas within a course. Clicking on a discussion displays the group ENA model and the associated codebook as shown in Figure \ref{fig:tool_group}. Instructors can edit their codebook by adding, removing, and editing keywords. They cannot edit the number of topics since~\cite{saravani2022automated} found 5 to be the optimal number of topics for grouping keywords. However, they can edit the names of the topics to be more descriptive since LDA will just assign the topics as numbers 0-4. Once the edits are completed, the ENA models automatically update to account for the codebook changes. Instructors are also able to view individual students' networks and the discussion posts that contributed to the network (Figure \ref{fig:tool_individual}). 

Shown in Figure \ref{fig:tool_group}, OnDiscuss provides a link to the discussion in Canvas and a link to the SpeedGrader for easy access to grade and view the original discussion in Canvas. There was also a link to the "ENA User Manual," which provided reinforcement of what ENA is and how to interpret an example network. A link to download the comma separated value (CSV) file was provided in the correct format for the ENA Web Tool~\cite{ena_web_tool}.

\subsection{Experimental Procedure}
The experiment was conducted as an empirical observational study~\cite{pyrczak2018}, where we observed the instructor's interaction with OnDiscuss during semi-structured interview~\cite{magaldi2020semi} sessions that lasted around one hour. We recorded the interview sessions and took notes about instructor's interactions and questions. We provided assistance in response to any inquiries raised by the instructor during the session. Because discussion topics were pulled from past semesters of the course, the instructors were asked to review the selected discussion prior to the interview session. This was to allow the instructors to reacquaint themselves with the discussion topic and the students' postings and discourse. 

To begin the session, we delivered a presentation that went over the basics of what ENA is, a case study illustrating ENA interpretation, and the impact of discussion data and codebook modifications on visualizations. The presentation began by explaining that ENA identifies the co-occurrences in segments of discourse data and modeling the weighted structure of co-occurrences as a dynamic network model~\cite{shaffer2017quantitative}. We then presented a case study on one discussion topic from a single semester of a course. An initial codebook was generated using LDA and was used to create the group class and individual student models where instructors learned that the nodes represented topics, the thickness of the lines represented the strength of connections between topics, and the other points in the group network represented the centroids of the individual students' networks. The presentation gave examples of student posts and how keywords co-occurred in the discourse data to create the ENA model. Because this case study was performed on a single discussion topic from a single semester, additional networks of the same discussion topic from a different semester were shown to demonstrate the impact of the discussion data using the same codebook. Finally, we made our own edits to the initial codebook and demonstrated how such modifications could influence the resulting networks. This presentation was delivered to both instructors, irrespective of their familiarity with ENA. 

After the presentation the instructors navigated to the chosen discussion assignment in OnDiscuss. The initial codebook was completely generated by LDA from previous semesters of the same discussion topic~\cite{saravani2022automated}. The instructors were allowed to make as many edits and iterations of the codebook while inspecting the models produced by each iteration. Once the instructor was satisfied with the codebook and/or the group and individual ENA models, a closing semi-structured interview was conducted.

\section{Results}
\subsection{Instructor A}
\begin{figure}[h]
    \centering
    \begin{subfigure}[b]{0.45\textwidth}
        \centering
        \includegraphics[width=1.2\textwidth]{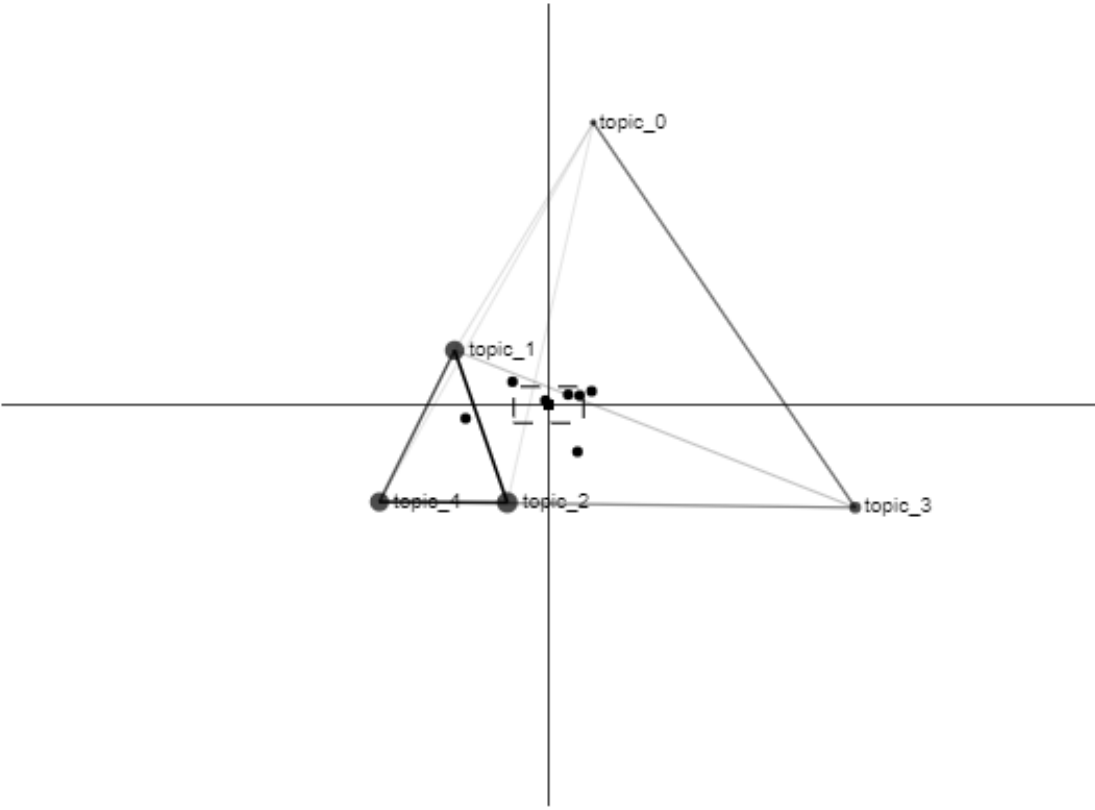}
        \caption{Initial Network}
        \label{fig:A_initial}
     \end{subfigure}
     \begin{subfigure}[b]{0.45\textwidth}
        \centering
        \includegraphics[width=1.2\textwidth]{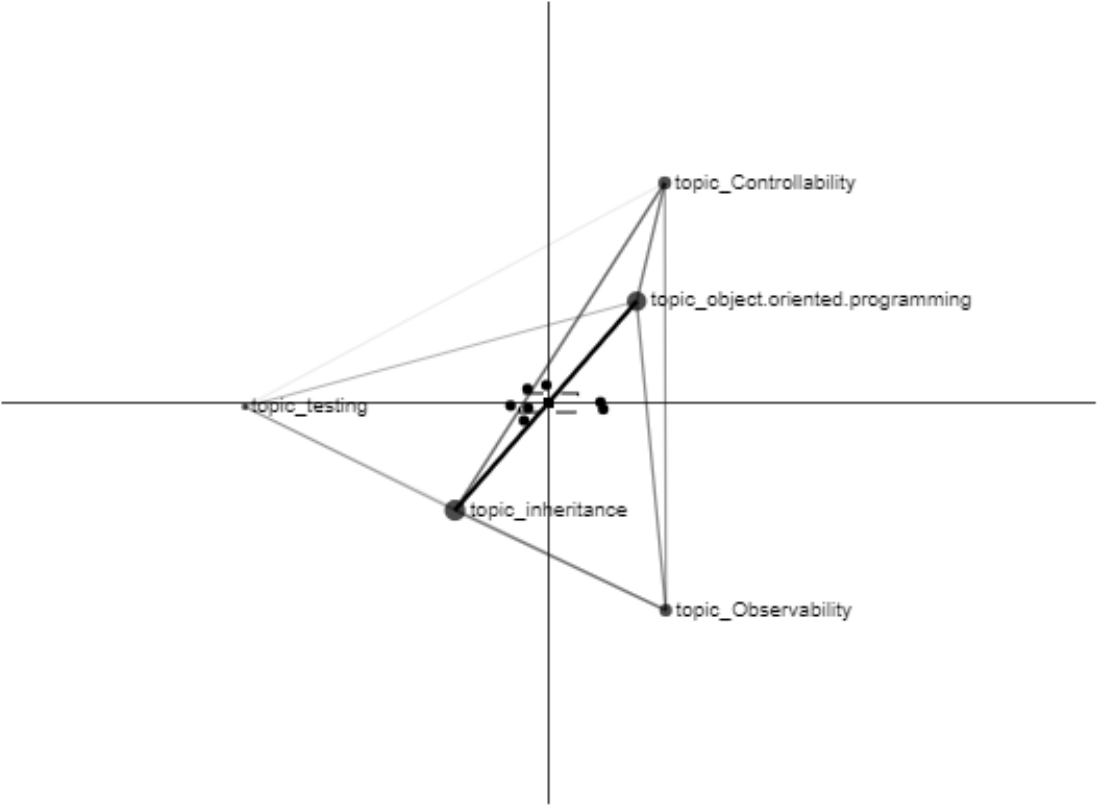}
        \caption{Best Network}
        \label{fig:A_best}
     \end{subfigure}
    \caption{Instructor A Group ENA Visualizations}
\end{figure}

The novice with ENA asked many questions about the basics of ENA during the presentation. For example, \textit{What is a codebook?}, \textit{How are the topics positioned in 2D space?}, and \textit{What should I be expecting to understand from these visualizations?}. After responding to the instructor's questions and completing the presentation, the instructor navigated to their chosen discussion on OnDiscuss. At this point, they had access to the initial codebook created from running LDA on all the previous semesters of the same discussion topic (Table \ref{tab:instructorA_initial_codebook}) the resulting group network (Figure \ref{fig:A_initial}), and all the individual networks. 

\begin{table}[h]
    \centering
    \setlength{\tabcolsep}{7pt} 
    \renewcommand{\arraystretch}{1.5}
    \resizebox{0.95\textwidth}{!}{
    \begin{minipage}{\textwidth}
    \caption{Instructor A Initial Codebook}
    \label{tab:instructorA_initial_codebook}
    \begin{tabular}{| m{1cm} | p{10cm}|}
    \hline
        \textbf{Topic} & \textbf{Keywords} \\
    \hline
        0 & devic, interfac, child, applic, potenti, post, input\_paramet, parent, behavior, run\\
    \hline
        1 & write, want, team, choic, custom, field, look, product, interfac, array\\
    \hline
        2 & boundari, import, select, api, handl, rest\_api, rest, encapsul, sure, partit\_test\\
    \hline
        3 & partit\_method, categori\_partit, leak, determin, memori\_leak, languag, applic, system, databas, partit\_test\\
    \hline
        4 & subclass, tester, abstract, output, group, model, oop, overlap, detect\_memori, disjoint\\
    \hline
    \end{tabular}
    \end{minipage}}
\end{table}

The instructor started their interaction with the tool by removing all the initial LDA generated keywords from each topic and added their own keywords. They said that half of all the automatically generated keywords were useful individually but the groupings of the keywords into topics was not helpful. Consequently, they removed all keywords and added those they thought would be present. The first iteration of their codebook did not have enough keywords that co-occurred in the discourse data so the network only consisted of a single line. 

We intervened and suggested that this could've been due to multiple reasons such as not including the stems of keywords, not choosing enough keywords to represent a topic, and choosing keywords not present in the discussion. Taking this advice, the instructor included the proper word stems, added more keywords to existing topics, and completely reworked a topic that only had 2 keywords to a new topic with 12 keywords. The codebook that they were most pleased with is shown in Table \ref{tab:instructorA_best_codebook} and the resulting group network is shown in Figure \ref{fig:A_best}.

\begin{table}[h]
    \centering
    \setlength{\tabcolsep}{7pt}
    \renewcommand{\arraystretch}{1.5}
    \resizebox{0.95\textwidth}{!}{\begin{minipage}{\textwidth}
    \caption{Instructor A Best Codebook}
    \label{tab:instructorA_best_codebook}
    \begin{tabular}{| m{2.5cm} | p{8.5cm}|}
    \hline
    \textbf{Code}  &  \textbf{Keywords} \\
    \hline
    Observability & observability, visible, get, state, visibility, observable, getter, access, field, accessor\\
    \hline
    Controllability & configure, object, control, modify, state, mutate, mutator, update\\
    \hline
    inheritance & inheritance, child class, parent class, overriding, depth, subclass, superclass, sub class, super class, inherit, interface, abstract class\\
    \hline
    testing & black box, black-box, white-box, white box, automate, automation, industry, difficult, easy\\
    \hline
    object oriented programming & public, private, protected, package, simple, complex, abstraction, specialization, data, encapsulation, method, field\\
    \hline
    \end{tabular}
    \end{minipage}}
\end{table}

\begin{figure}[h]
    \centering
    \begin{subfigure}[b]{0.45\textwidth}
        \centering
        \includegraphics[width=1.2\textwidth]{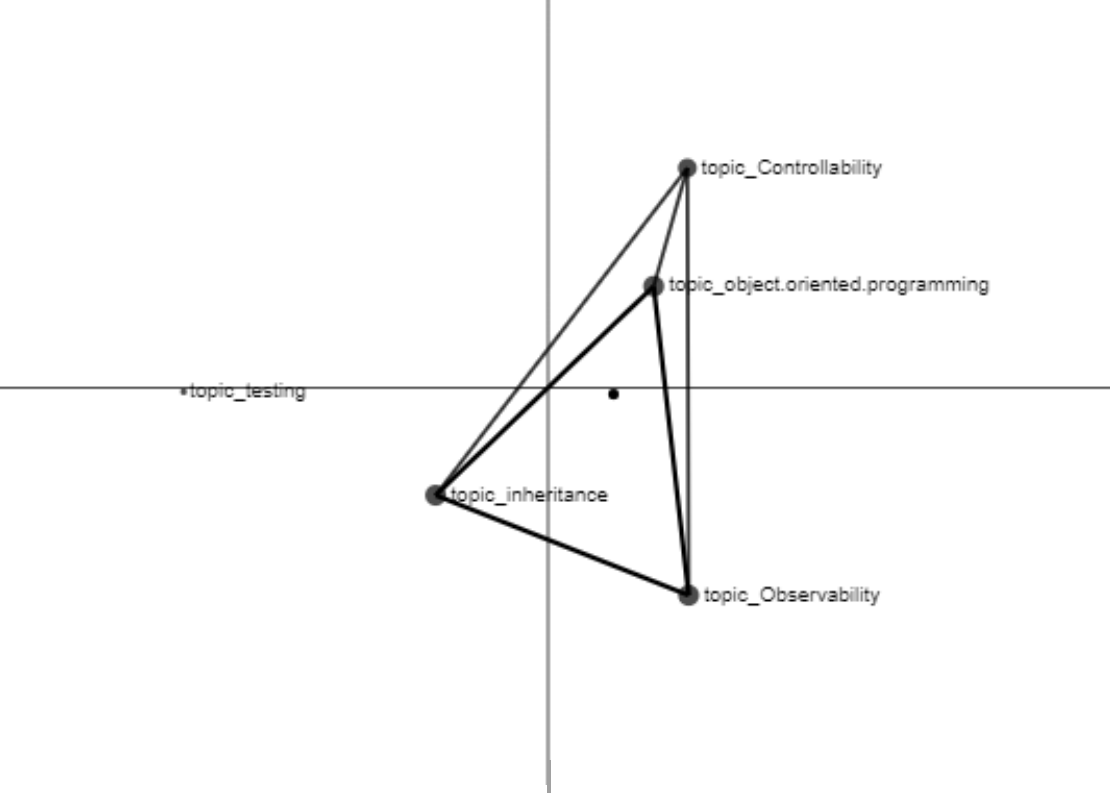}
        \caption{Strong Student Network}
        \label{fig:strong_student}
     \end{subfigure}
     \begin{subfigure}[b]{0.45\textwidth}
        \centering
        \includegraphics[width=1.2\textwidth]{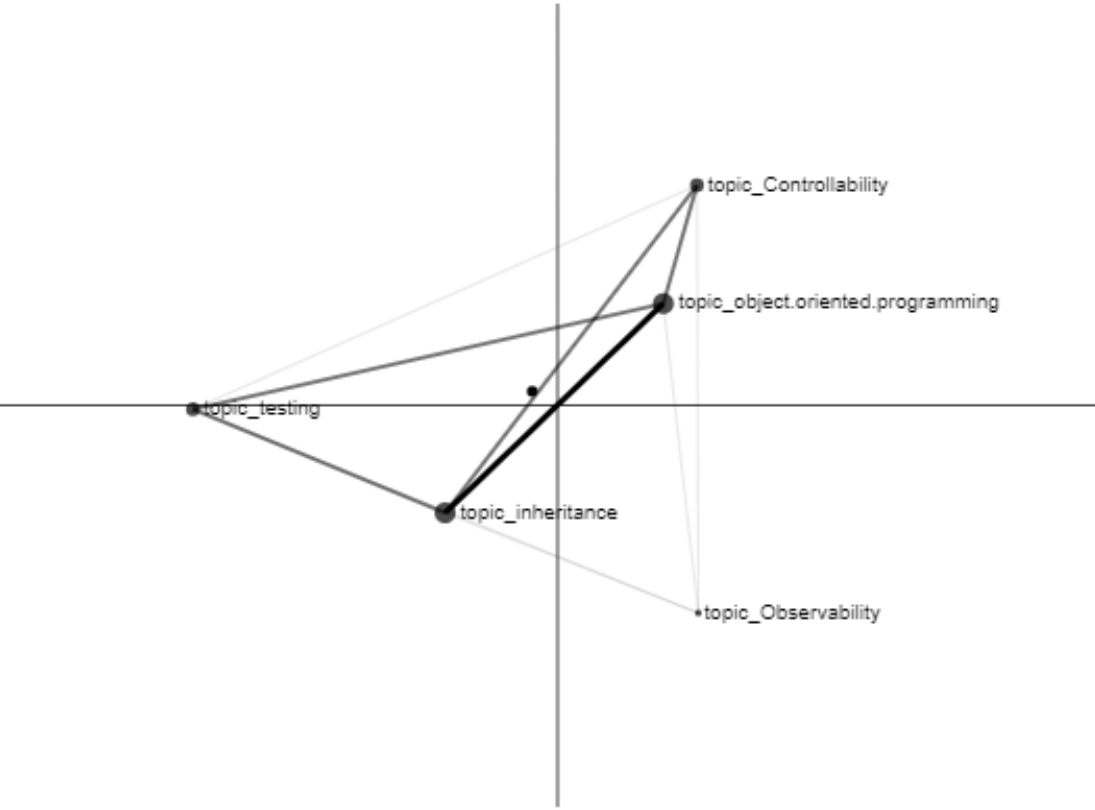}
        \caption{Weak Student Network}
        \label{fig:weak_student}
     \end{subfigure}
    \caption{Instructor A Individual ENA Visualizations}
\end{figure}

Analyzing the networks from the codebook shown in Table \ref{tab:instructorA_best_codebook}, Instructor A stated that the connection strength between topics in the group ENA shown in Figure \ref{fig:A_best} did indeed reinforce what they had gathered from rereading the discussion. The individual ENA graphs were also representative of both the "strong" and "weak" students' contributions. When analyzing a strong student's network (Figure \ref{fig:strong_student}), the instructor was shocked to not see any connections to the topic "testing" but stated, \textit{This student made good strong connections to controllability and observability}. When analyzing a weak student's network (Figure \ref{fig:weak_student}), the instructor stated, \textit{I was expecting weak connections to controllability and observability because they were initially answering a different question from the prompt}. 

When asked if having these visualizations would impact how they reframe future discussion prompts, Instructor A stated they wouldn't change the prompt but rather better prepare the students in class for the discussion. Instructor A explained that they'd display the group network to the entire class to discuss the topics without connections. The instructor saw this as a helpful tool to then reinforce concepts that were missed without needing to read the entire discussion thus saving time and energy. They also noted that these ENA models would not only be helpful to the instructor but also to the students.

\subsection{Instructor B}
During the presentation, the expert with ENA asked the following questions: \textit{What is LDA?}, \textit{What stanza is being used?}, and \textit{How are the student points being placed?}. When Instructor B began exploring OnDiscuss they were pleased with the initial codebook provided (Table \ref{tab:instructorB_initial_codebook}). This initial codebook was created with more discussion data from various semesters so LDA was able to perform better topic modeling. Since this codebook was used in previous works, the topics had descriptive names instead of the LDA assigned numbers and it had some added keywords from the same instructor~\cite{moraes2023combining}.

\begin{table}[h]
    \centering
    \setlength{\tabcolsep}{7pt} 
    \renewcommand{\arraystretch}{1.5}
    \resizebox{0.95\textwidth}{!}{
    \begin{minipage}{\textwidth}
    \caption{Instructor B Initial Codebook}
    \label{tab:instructorB_initial_codebook}
    \begin{tabular}{| m{2.5cm} | p{8.5cm}|}
    \hline
        \textbf{Topic} & \textbf{Keywords} \\
    \hline
        effortful learning & desire, plf, resonate, parachute, land, jump, commun, parachute land, land fall, difficult, difficulties, mistakes, failure, effortful learning, desirable difficulty, desirable, effortful\\
    \hline
        beyond learning styles & dylexia, learn style, individual, learn differ, disable, intelligent, prefer, support, dyslex, focus, instructional style, learning styles\\
    \hline
        illusion of mastery & confidence, feedback, calibration, confidence memory, accuracy, peer, answer, event, state, calibration learn, illusion of mastery, illusions of mastery, misunderstanding, illusion of knowing, illusions of knowing, illusion of learning, illusions of learning, re read, cram\\
    \hline
        retrieval practice spaced out practice interleaving & mass, mass practice, interleaving practice, space retrieval, tend, day, long term, week, myth, practice space, retrieval practice, retrieval process, testing effect, test effect, recall knowledge, retrieval, actively retrieving, periodically testing, retrieval activity, retrieval activities, low stakes, effective retrieval must be repeated, flash cards, quizzing, practice and retrieval, quiz over time, continually retrieve the information, frequently quizzing, retrieval practice activity, retrieval practice activities, testing efforts, active retrieval, practice, testing for its benefit in the learning process, short quiz, active recall, process of retrieval, practice sessions, self testing, recall the information, RPA, RPAs, spacing out, spacing out practice, spaced practice, spacing practice, spaced out practice, spaced out, spaced retrieval, space retrieval, space practice, retrieval spaced, retrieve spaced, spaced application, spaced knowledge, space knowledge, spaced retrieval, retrieval practice is spaced, interleaving, interleaved practice, interleave, interleaved\\
    \hline
    \end{tabular}
    \end{minipage}}
\end{table}

\begin{figure}[h]
    \centering
    \begin{subfigure}[b]{0.45\textwidth}
        \centering
        \includegraphics[width=1.1\textwidth]{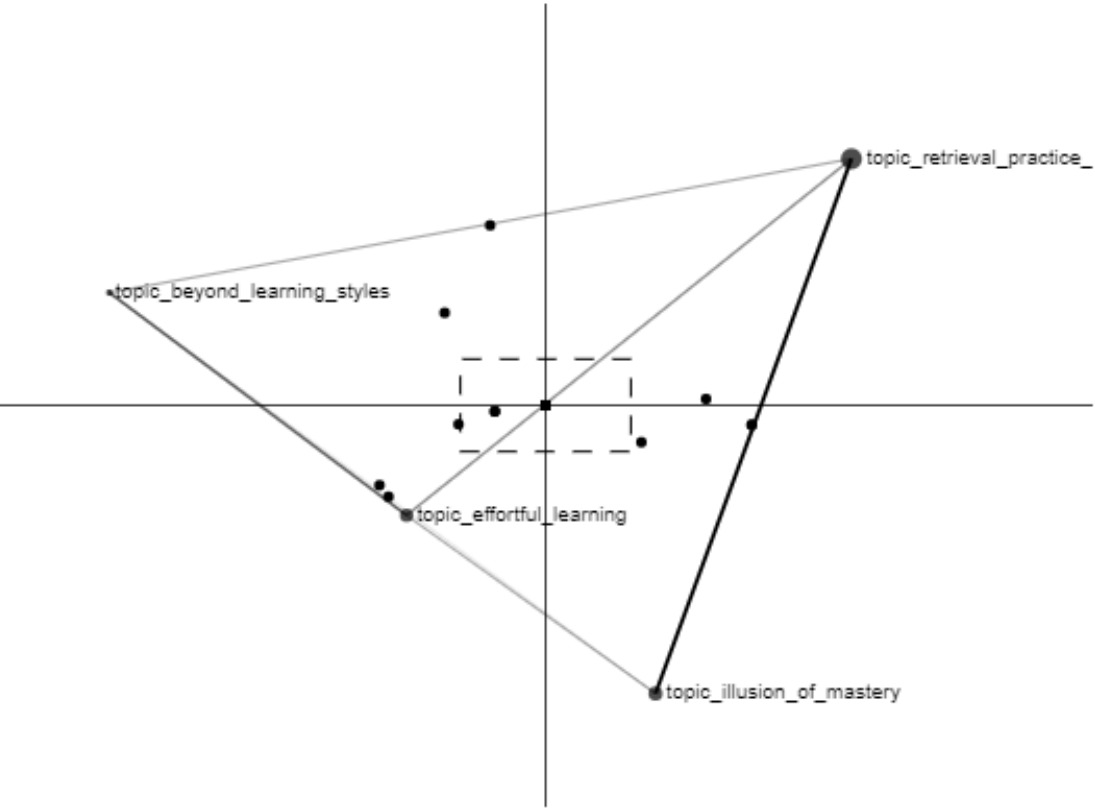}
        \caption{Initial Network}
        \label{fig:B_initial}
     \end{subfigure}
     \begin{subfigure}[b]{0.45\textwidth}
        \centering
        \includegraphics[width=1.1\textwidth]{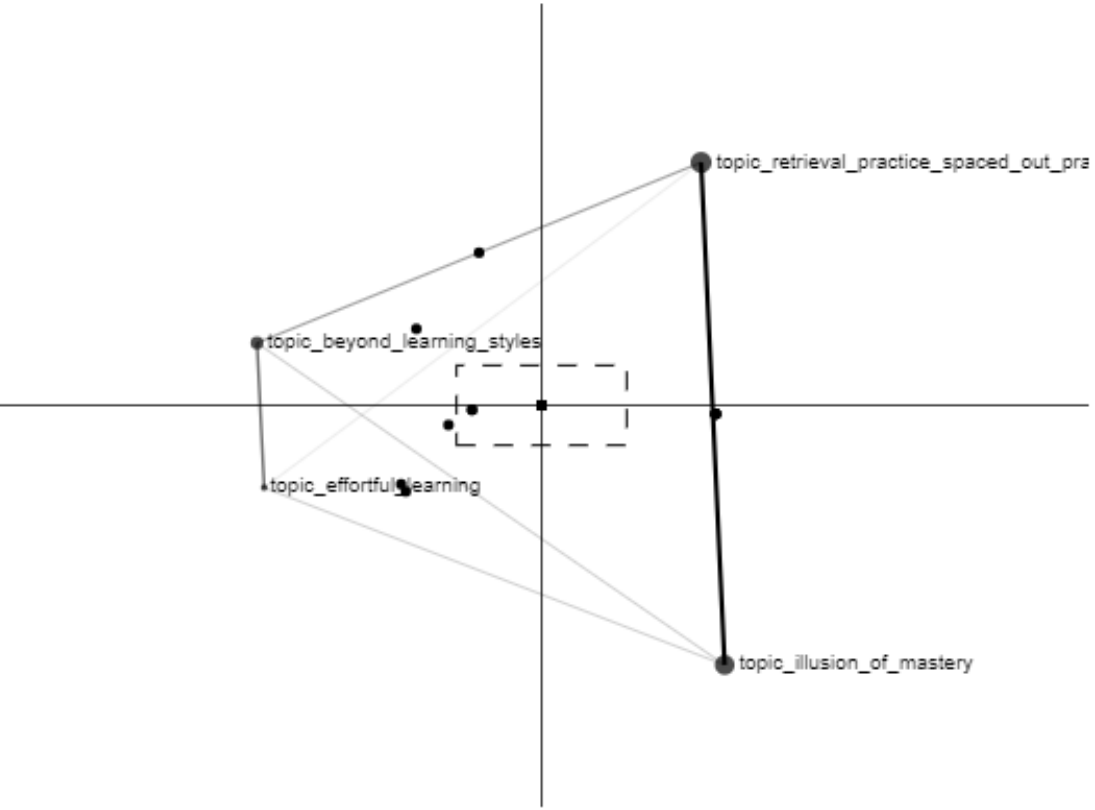}
        \caption{Best Network}
        \label{fig:B_best}
     \end{subfigure}
    \caption{Instructor B Group ENA Visualizations}
\end{figure}

This instructor made very minimal edits to the initial codebook. They only removed "fall" and "desire difficulty" from the "effortful learning" topic. The network model changed as the instructor expected; however, they didn't think much could be concluded from the change between Figure \ref{fig:B_initial} and Figure \ref{fig:B_best}. Figure \ref{fig:B_best} had a new connection between "beyond learning styles" and "illusion of mastery" that did not appear in Figure \ref{fig:B_initial}. Comparing networks between codebook edits wasn't meaningful to Instructor B. Instead, they wanted a baseline model to compare the resulting network to in order to make more informed edits to the codebook. Instructor B suggested creating a baseline model from a textbook or an asynchronous online discussion amongst the instructors responding to the same discussion prompt as the students.

While examining individual student interactions, the instructor expressed the need for two separate networks: one focusing solely on each student's initial post and another incorporating both their original post and subsequent replies. This would provide a more isolated insight into the students' contribution and engagement with other students as well as their initial contribution. Additionally, the instructor suggested that a clearer indication of occurring keywords would be helpful when reading the student's discussion posts in the tool rather than just the plain text.

\section{Discussion}

\subsection{RQ1. Are ENA visualizations of asynchronous online discussion data helpful for instructors who are novices in ENA?}

The novice instructor started with no understanding of ENA. Throughout the presentation and the time exploring the tool, this instructor was unsure what made a codebook better in terms of what made a topic a correct grouping of keywords. They were also unsure what makes an ENA network better in terms of the number of nodes, number and thickness of the edges, and placement of the centroid. Since this instructor is in the field of computer science, they seemed more inclined to understand the math behind the ENA visualizations. Future work should be done with novice instructors in other fields to determine if they also have the same inclination. This could inform future revisions to the presentation to emphasize and include more of the mathematical background behind ENA.

Once the instructor created a codebook with keywords they were pleased with, they were able to successfully draw conclusions from the networks. The placement of the points in the ENA network was still confusing to this instructor; however, they were able to extract lots of meaning from the thickness of the edges of both the individual and group networks. They were able to make interpretations such as, \textit{The entire class isn't making many connections to these topics so I should emphasize them more in lectures} and \textit{I thought this student made a strong contribution to the discussion and it's nice to see that visually they made lots of strong connections to all of the topics}. Instructor A stated that once they understood how to create a better codebook they'd feel even more familiar and inclined to use ENA.

The ENA presentation likely played a vital role in the novice instructor's basic understanding of ENA which manifested into their own ENA interpretations. We hypothesize that additional exposure to modifying codebooks and interpreting ENA networks from other discussions would be beneficial for the instructor's continued development. However, through just a single session this instructor demonstrated promising proficiency in ENA, indicating the potential for other instructors unfamiliar with ENA. 

\subsection{RQ2. Do the ENA visualizations have the potential to reduce the time and effort spent assessing asynchronous online discussions for instructors of any experience level with ENA?}

Both Instructor A and B reported that having these ENA networks and a modifiable codebook was a helpful supplemental tool for evaluating their asynchronous online discussions. Both said in the semi-structured interview that these supplemental networks would make grading and analyzing the discussions faster. This tool is not meant to fully replace reading through the students' discussion posts but is instead a supplemental tool. The instructors stated it was faster for them to assess the discussions with the visual aid provided by the ENA networks. The instructors also stated by utilizing the individual networks they'd be able to provide more personalized feedback to individual students.

Instructor A found the group and individual networks helpful on their own, while Instructor B thought they were somewhat useful and needed more context, such as a baseline network, to be even more useful. Despite their differing perspectives and experiences with ENA, both instructors were able to draw their own conclusions from the networks based on their own codebooks. This diversity in interpretations highlights the power and flexibility of ENA and OnDiscuss, as instructors not only have the ability to handcraft their codebooks but also to derive nuanced insights tailored to their specific teaching contexts.

\subsection{Limitations}
The ENA networks only represent the raw text data within the discussions. Consequently, this may overlook other forms of media, such as images, videos, and links to external websites, which could provide valuable insights or perspectives that are not captured through text alone.

Another potential limitation concerns the study population. The instructors were purposefully chosen by us to participate in the research project. While this approach was necessary to ensure access to relevant discussion data and expertise with ENA, it introduced a selection bias. Also although the participants in our study are drawn from the same university, it is important to note that they exclusively represent graduate level courses with relatively small class sizes in two different disciplines: computer science and education.

Instructor A had considerably less available discussion data than Instructor B to create the codebook. Instructor A had 363 posts with 37,254 total words, while Instructor B had 2,648 posts with 444,364 total words. Because Instructor B had more data, LDA was able to provide more accurate topic modeling~\cite{blei2003latent}. Moreover, given that Instructor B's codebook has been utilized in previous studies, it is plausible that the codebook may have undergone refinements over time, potentially rendering it more polished compared to Instructor A's. 

\section{Conclusion and Future Works}
In this study, we explored the potential of utilizing OnDiscuss, a tool that enables instructors to edit their own codebook and visualize the resulting ENA visualizations in real time. By providing instructors with the ability to directly interact with and customize, OnDiscuss opens up new opportunities to popularize ENA and make it more accessible to educators who may or may not be familiar with the intricacies of ENA.

Our findings highlight several key insights regarding the implications of this tool for enhancing the accessibility and usability of ENA as a learning analytics visualization tool. OnDiscuss is helpful to those unfamiliar with ENA since it abstracts many of the intricacies of ENA by providing an easy interface to manipulate a codebook and thus the resulting ENA networks. Future refinements, such as the addition of a baseline ENA model, can make it more helpful to those familiar with ENA. Despite the tool's automated keyword generation capabilities, it is clear that instructor intervention remains crucial for refining the codebook. Therefore, while automated techniques like Latent Dirichlet Allocation (LDA) provide valuable insights given a large amount of data, they must be complemented by expert guidance.

Many potential future works could stem from this study in terms of both adding new features to the tool and exploring different populations. Adding a baseline ENA model from either the textbook or discussion posts of the instructor could have potential to assist the instructor even further by allowing them to make comparisons between the class and base model. Future research should also explore applying OnDiscuss to different subjects other than Computer Science and Education, to undergraduate courses, and to even larger class sizes since it would be even more time consuming to assess the discussion assignments. This study was carried out on historical discussion data, underscoring the necessity for future studies to be done on an in progress semester of discussions to uncover further insights into the implications of utilizing OnDiscuss.

\bibliographystyle{splncs04}
\bibliography{references}

\end{document}